\pdfoutput=1
\documentclass[acmtog]{acmart}

\usepackage{booktabs} 
\usepackage{subfigure}

\usepackage[ruled]{algorithm2e} 

\SetAlFnt{\small}
\SetAlCapFnt{\small}
\SetAlCapNameFnt{\small}
\SetAlCapHSkip{0pt}

\begin{document}
\title{Style-compatible Object Recommendation for Multi-room Indoor Scene Synthesis}

\author{Yu He}
\affiliation{%
  \institution{Tsinghua University}
  \city{Beijing}
  \country{China}}
\email{hooyeeevan@tsinghua.edu.cn}

\author{Yun Cai}
\affiliation{%
  \institution{Tsinghua University}
  \city{Beijing}
  \country{China}}
\email{caiy18@mails.tsinghua.edu.cn}

\author{Yuanchen Guo}
\affiliation{%
  \institution{Tsinghua University}
  \city{Beijing}
  \country{China}}
\email{guoyc19@mails.tsinghua.edu.cn}

\author{Zhengning Liu}
\affiliation{%
  \institution{Tsinghua University}
  \city{Beijing}
  \country{China}}
\email{lzhengning@gmail.com}

\author{Shaokui Zhang}
\affiliation{%
  \institution{Tsinghua University}
  \city{Beijing}
  \country{China}}
\email{zhangsk18@mails.tsinghua.edu.cn}

\author{Songhai Zhang}
\affiliation{%
  \institution{Tsinghua University}
  \city{Beijing}
  \country{China}}
\email{shz@tsinghua.edu.cn}

\author{Hongbo Fu}
\affiliation{%
  \institution{City University of Hong Kong}
  \city{Hongkong}
  \country{China}}
\email{fuplus@gmail.com}

\author{Shengyong Chen}
\affiliation{%
  \institution{Tianjin University of Technology}
  \city{Tianjin}
  \country{China}}
\email{sy@ieee.org}


\begin{abstract}
Traditional indoor scene synthesis methods often take a two-step approach: object selection and object arrangement. Current state-of-the-art object selection approaches are based on convolutional neural networks (CNNs) and can produce realistic scenes for a single room. However, they cannot be directly extended to synthesize style-compatible scenes for multiple rooms with different functions. To address this issue, we treat the object selection problem as combinatorial optimization based on a Labeled LDA (L-LDA) model. We first calculate occurrence probability distribution of object categories according to a topic model, and then sample objects from each category considering their function diversity along with style compatibility, while regarding not only separate rooms, but also associations among rooms. User study shows that our method outperforms the baselines by incorporating multi-function and multi-room settings with style constraints, and sometimes even produces plausible scenes comparable to those produced by professional designers.
\end{abstract}

%
%


%
%


\begin{teaserfigure}
	\centering
	\includegraphics[width=7in]{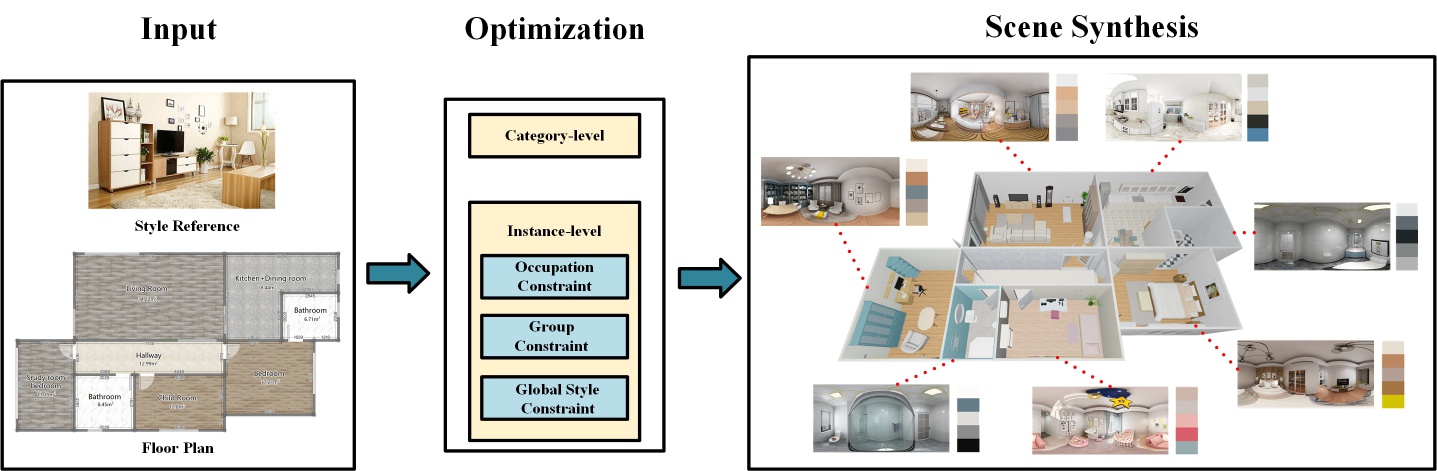}
	\caption{We present an object recommendation approach for style-compatible multi-room indoor scene synthesis. Given a labeled floorplan and a style reference image (left box), we assign a template paranoma image which has similar style with the reference image for each room based on the room type (right box), then perform object instance selection through category-level and instance-level recommendation (middle box). We use an existing method for object arrangement and the generated result is shown in the right box.}
	\label{pipeline}
\end{teaserfigure}

\maketitle

\section{Introduction}


Automatic indoor scene synthesis benefits various applications such as interior design and virtual reality. Existing indoor scene synthesis methods have mainly focused on object selection and arrangement, that is, what objects to choose and where to put them. Some recent works utilize deep convolutional neural networks (CNNs) to model object configurations by iteratively analyzing top-down images of indoor scenes. However, when it comes to practical interior design, there are three obvious drawbacks of existing approaches. First, they are all targeted for single rooms instead of multiple rooms in a residence. If multiple rooms are processed individually one by one, it could happen that rooms of the same types share similar sets of object categories, which we call ``function collapse''. Second, it often requires a single room type for each room. However, it is common that a single room might need to serve for multiple purposes due to the space constraints. For example, a dining table might have to be placed in a living room due to the lack of a dining room. Third, style compatibility between objects is often ignored, though styles are crucial in order to generate plausible scenes. \cite{chen2015magic} assign textures and materials for each object instance in a given 3D indoor scene using a data-driven method. However, their method focuses on material suggestion but not scene synthesis, and is more difficult to handle a multi-room scenario, since there exists some style compatibility for object instances in different rooms that have the same categories.

To address the above issues, we propose a practical indoor scene synthesis pipeline, which performs object recommendation and optimizes style compatibility not only in a multi-functional room, but also among multiple rooms in a residence. Specifically, our method automatically synthesize indoor scenes, given as input a residential floor plan with one or more labels assigned to each room according to their function needs, and an image of an existing scene as a style template. Regarding each room as a multi-labeled document and objects inside it as words, our method fits a topic model using Labeled Latent Dirichlet Allocation (Labeled LDA, L-LDA) \cite{ramage2009labeled} from the room data of the SUNCG dataset \cite{song2017semantic}. For each of the user-labeled empty rooms, we generate occurrence probability of each object category from the above topic model. Then for each room, we compute a style palette based on the user-provided style template, and select object instances by solving a combinatorial optimization, subject to multiple constraints related to space area, funcion diversity and style harmony.
We improve the model proposed in \cite{chen2015magic} to achieve style compatibility across multiple rooms. Once we have all the object instances chosen for each room, we use an existing object arrangement method \cite{shaokui} to synthesize a well-arranged indoor scene.

To show the instance diversity brought by our multi-label and multi-room settings, we compare our synthesis results with those generated separately by existing single-label single-room methods. For style compatibility, we compare our stylized results with material suggestions given by \cite{chen2015magic} and show that our method performs better in both room-level and residence-level. A user study shows that our method has the ability to generate residence-level indoor designs comparable to those produced by professional interior designers.

In summary, the contributions of our paper are as follows: 
\begin{itemize}
    \item We present a style-compatible multi-room indoor scene synthesis pipeline, which can generate practical interior scenes comparable to those created by professional interior designers.
    \item Our work is the first to incorporate style compatibility for indoor scene synthesis. Our model automatically selects object instances with respect to a user-provided style template.
    \item We are the first to explore residence-level planning for indoor scene synthesis. Our model can select object instances considering associations among rooms instead of treating each room separately.
\end{itemize}

\section{Related Work}
\subsection{Indoor Scene Synthesis}

Indoor scene synthesis has been studied for many years. Early solutions adopt interior design guidelines of humans as constraints to generate reasonable layouts  \cite{merrell2011interactive}. With the emerging of large indoor scene datasets, prior knowledge encoded in the massive data has been converted to design rules for synthesizing scene layouts \cite{weiss2018fast}. The statistical relationships between different objects learned from examples are also applied to object arrangement \cite{yu2011make}. Fisher et al. \shortcite{fisher2012example} proposed the first data-driven scene synthesis method, using a directional graphical model of object co-occurrence and a Gaussian Mixture Model (GMM) of pairwise spatial relationships to learn prior knowledge of object occurrence and placement. Other learned priors include human-centric stochastic grammars \cite{qi2018human}, priors based on activity-object relationship graphs \cite{fu2017adaptive}, topic model \cite{liang2017automatic}, layout patterns based on directed graphs incorporated with a GMM \cite{henderson2017generative}, and discrete priors extracted by tests for complete spatial randomness (CSR) \cite{shaokui}. However, the above methods are all designed for indoor scene synthesis of a single-label room. There is yet no solution of scene synthesis for multi-label rooms or multiple rooms. In this paper, we focus on object recommendation for indoor scene synthesis given a floor plan with multiple rooms, each of which might be multi-functional.

\subsection{Supervised Topic Model}

Supervised topic models provide a powerful tool for latent data discovery. Supervised Latent Dirichlet Allocation (SLDA) \cite{mcauliffe2008supervised} and its variations are the most popular methods of supervised topic model. SLDA supposes that the labels are generated based on a mixture of empirical topics for each document, and limits one document to have only one label. A similar work named Discriminatively Trained LDA (DiscLDA) \cite{lacoste2009disclda} cannot handle the case of multi-label documents as well. Multi-Multinomial LDA (MM-LDA) models each document as a bag of words with a bag of labels \cite{ramage2009clustering}, and derives the topic for each observation from shared topic distribution. MM-LDA is not constrained to single-label documents, but the learned topics cannot directly correspond to the labels in the label set. Another multi-label document topic extraction model is Labeled LDA (L-LDA) \cite{ramage2009labeled}, which presents a solution to the credit attribution problem \cite{mcauliffe2008supervised}. In this paper, we treat the SUNCG dataset as a corpus consisting of multi-label documents, and utilize L-LDA to extract the latent relationship between object categories and room types. We apply this latent topic relationship to object instance selection for creating indoor scenes.

\subsection{Style Compatibility}

Style compatibility has been studied in two aspects: color compatibility and appearance compatibility. A considerable amount of effort has been devoted to color compatibility studies. Various methods have been proposed in evaluating the compatibility of multi-color mixing  \cite{o2011color,cohen2006color}. A color harmony template was developed to adjust the color matching rate of images \cite{cohen2006color}. O'Donovan et al. \shortcite{o2011color} proposed a data-driven 5-color palette rating model to evaluate the color compatibility of images.
Appearance compatibility is determined by the material and texture of instance. Chajdas et al. \shortcite{chajdas2010assisted} proposed a method to automatically propagate the user-assigned textures to different surfaces in large scenes based on similarity measures. Leifman and Tal \shortcite{leifman2012mesh} presented a mesh colorization method, which requires users to specify initial colors for certain areas, and propagates these colors to surrounding meshes according to intensity similarity. Nguyen et al. \shortcite{nguyen20123d} proposed a method for obtaining material styles from a single example image via heuristic estimation, and then propagating the obtained styles to target 3D scenes. 

With the emerging of large databases of object styles like \cite{bell2013opensurfaces}, data-driven approaches for style suggestion have become a mainstream. However, there's very few work for the automatic assignment of materials and textures for indoor scenes. \cite{jain2012material} used a 3D database to train a factor graph containing material and shape contexts, and materials can be automatically assigned to 3D models based on the factor graph. A system called Magic Decorator is proposed in \cite{chen2015magic} which casts the problem of material suggestion as a combinatorial optimization. In this paper, we extend their approach to perform automatic material and texture suggestion in object recommendation.

\section{Overview}

The input to our method is a floor plan, with one or more labels assigned to each room, and an example image for style template. Our goal is to automatically generate indoor scenes that meet both functional and aesthetic requirements as shown in Figure \ref{pipeline}. Since this paper only focuses on object recommendation, we adopt an existing approach on object layout for each room respectively once we get the selected object instances. We propose a coarse-to-fine strategy for selection because object categories determine functionality and specific instances determine appearance diversity, so it's natural to select category and instance in separate steps.

\paragraph{Category-level recommendation} Given a floor plan with each room labeled with one or more room types, category-level recommendation gives what object categories appear in each room along with the number of instances for each category. Regarding each labeled room as a bag of objects, it's natural to apply a topic model for relevance analysis between object categories and room types of the dataset. Here we fit a widely-used L-LDA model on our modified SUNCG dataset S-SUNCG, and this will give us object category occurrence probability of each room type. For multi-labeled rooms, we simply add up the probability for each category respectively. Results show that this simple strategy performs well. If there're multiple rooms of the same category, our motivation is that there should be diversity for objects of the same function, single bed and double bed for example. Since SUNCG dataset only have very coarse categorization (e.g. beds), we achieve this by annotating objects with more detailed category (e.g. single beds and double beds) in our S-SUNCG dataset and propose a category list split method to assign different category list for rooms with the same label. Experiments show that our method effectively solves the function collapse problem and further generates better synthesized results. We give the number of instances for each recommended category in each room based on the statistical results of the dataset, since the number pattern of the objects can be easily revealed through a simple counting process.

\paragraph{Instance-level recommendation} Given the occurrence probability and instance number of object categories of each room, along with a style template image, instance-level recommendation gives specific object instances. For each room, we first iteratively select object category instances of it according to a) the category's occurrence probability b) the number of instances needed to be selected from the category c) empty space left for the room. The iteration stops when there's no space left for any reasonable object instance. This selection strategy tend to have more dense layout than the existing methods, which is more realistic and aesthetic. Then, we automatically assign a style template image selected from a collections of style templates for each room based on the given image, and extract 5-color palette for them as style guidance. Under the constraints of these color palettes, we use a method motivated by \cite{chen2015magic} to apply styles for object instances from their available textures and materials. Different from \cite{chen2015magic}, we introduce the concept of object group to enhance intra-room style compatibility and also consider object connections among rooms for residential-level style compatibility.

\section{Dataset}

\begin{figure}[ht]
	\centering
	\includegraphics[width=3in]{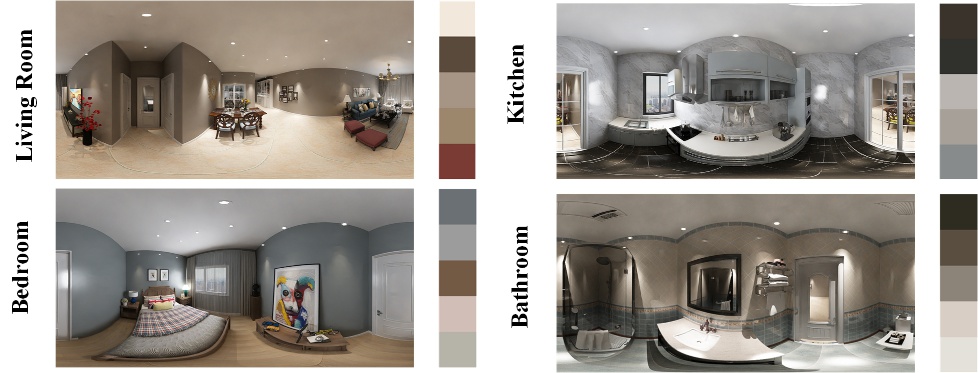}
	\caption{5-color palette extracted from panorama images of living room, bedroom, kitchen and bathroom in Structured3D dataset.}
	\label{panorama-palette}
\end{figure}

\begin{figure}[ht]
	\centering
	\includegraphics[width=3in]{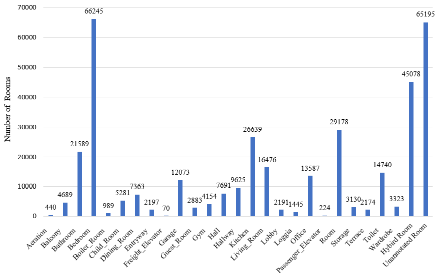}
	\caption{Room type statistics of original SUNCG dataset.}
	\label{room-type-statistic}
\end{figure}

In this paper, we utilize two large datasets, Structured3D \cite{zheng2019structured3d} and a modified version of SUNCG \cite{song2017semantic} dataset named S-SUNCG. We use rendered panorama images of indoor scenes  in Structured 3D as style template images and extract a 5-color palette for each of them. Figure \ref{panorama-palette} shows four panorama images with their extracted color palette. Details are in \cite{o2011color}. We made three major enhancement to the original SUNCG dataset, described as follows.
The original SUNCG dataset contains 45622 designed house layouts with a total of 368669 rooms. We can see from the statistical results in Figure \ref{room-type-statistic} that a considerable number of rooms (65195 of 368669) lack room type annotation, which is a waste of data.
\begin{figure}[ht]
	\centering
	\includegraphics[width=3in]{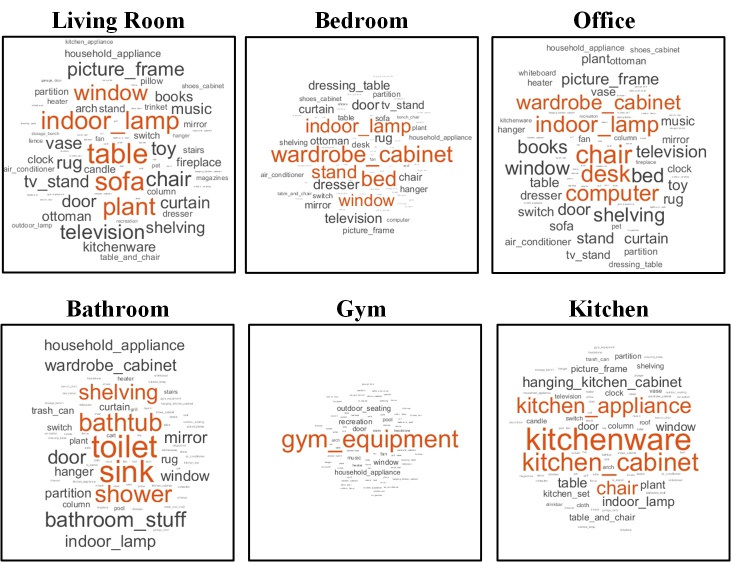}
	\caption{The occurrence probabilities calculated by our topic model of object categories in six different types of rooms (larger font size indicates larger probability).}
	\label{occ-room-category}
\end{figure}
\subsection{Room type prediction for un-annotated rooms}\label{room type}
We propose a room type prediction method based on the topic model. Treating the dataset as a corpus, each room $r$ is represented by a tuple consisting of a list of object category indices $ O^{\left ( r \right )}= (  o_{1},\cdots ,o_{N_{r}})$ and a list of binary room type indicators $\Lambda ^{\left ( r \right )}=\left ( l_{1},\cdots ,l_{U} \right )$ where each $o_{i}\in \left \{ 1,\cdots ,V \right \}$ and $l_{i}\in \left \{ 0,1 \right \}$. Here $N_{r}$ is the number of object categories in the room, $V$ is the number of all object categories and $U$ is the total number of unique room types in SUNCG dataset. Following the process in \cite{ramage2009labeled}, we first learn a topic model from the data, then predict room types for un-annotated rooms. 

In the learning stage, we draw a multinomial mixture distribution $\theta ^{r}$ over all $U$ room types for room $r$ from a Dirichlet prior $\alpha $, and a multinomial room type distributions over object category library $O$ for each room type $l_{u}$, from a Dirichlet prior $\beta$. The constraint that the room type prior $\alpha ^{(r)}$ is restricted to the set of annotated room types $R_{T}^{r}$. The joint probability of object and room type $p(O,R_{T}|\alpha ,\beta )$ can be learned from the training set. Figure \ref{occ-room-category} illustrates six examples of $p(O|R_{T})$ in word cloud, in which the occurrence probability of an instance category is represented by corresponding name tag size.

\begin{figure}[ht]
	\centering
	\includegraphics[width=3in]{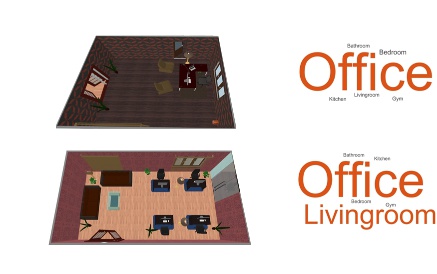}
	\caption{Two examples of automatic room type labeling using L-LDA.}
	\label{room-type-labeling}
\end{figure}

\begin{figure}[ht]
	\centering
	\includegraphics[width=3in]{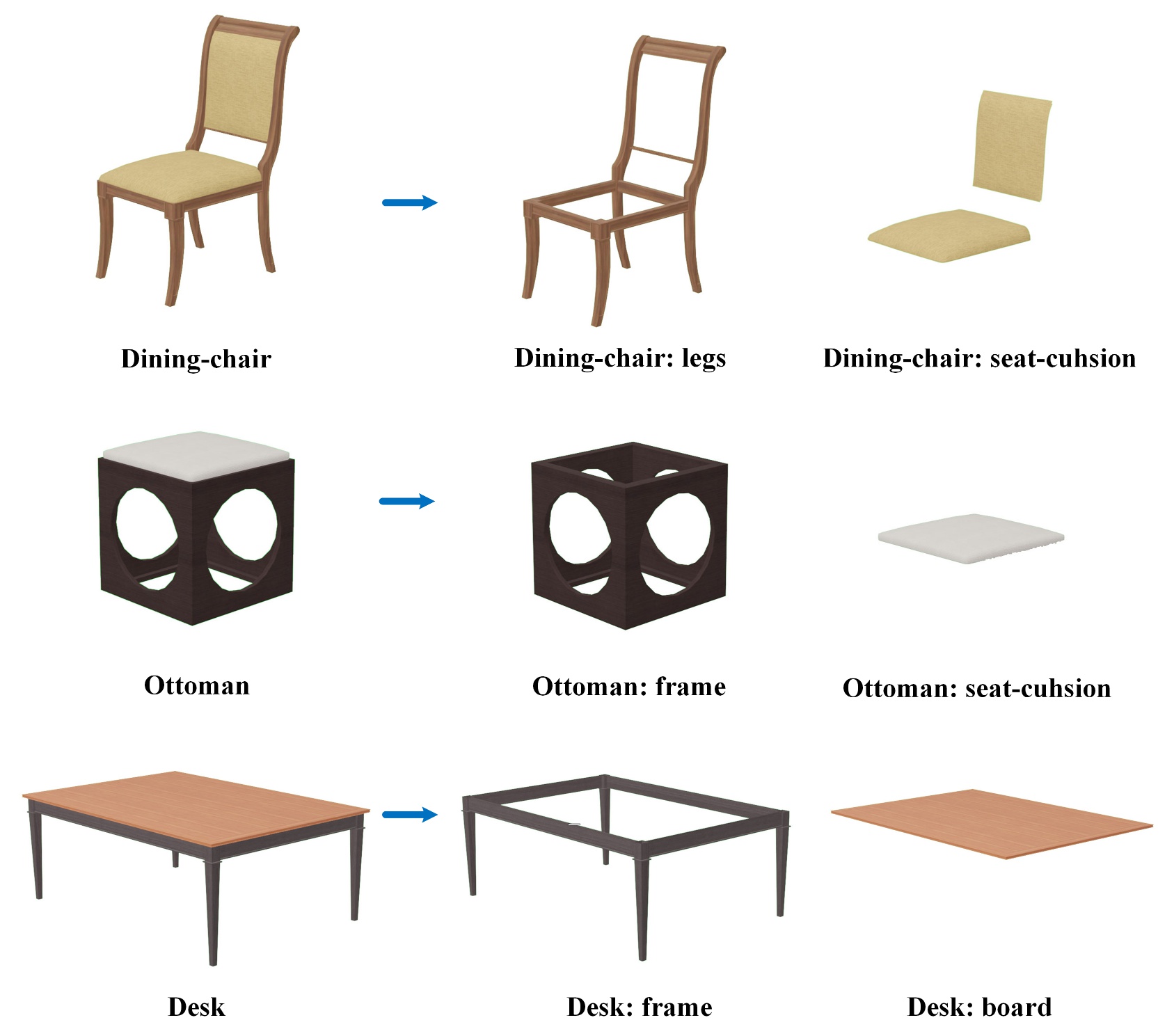}
	\caption{Model part labeling and material assignment of three example objects.}
	\label{model-part-labeling}
\end{figure}

In the prediction stage, the un-annotated rooms' most likely annotations can be inferred by properly thresholding its posterior probability over all room types. Figure \ref{room-type-labeling} gives two examples of the automatic annotation results. The left column shows two un-annotated room scenes in SUNCG dataset, and the right column shows the corresponding room types calculated by our topic model.

\subsection{Fine-grained labeling of object categories}
The original SUNCG dataset contains 2549 object models with 67072 material textures in 68 object categories. However, the object categories are ambiguous as shown in Figure \ref{occ-room-category} (e.g. kitchen$\_$appliance, household$\_$appliance), which cannot reflect the detailed functional attributes of the object and don’t provide enough information for object category recommendation. We re-classify the models into 165 new categories, which are more specified than the old ones. 

\subsection{Labeling of object parts}

Object instances are consisted of parts in the SUNCG dataset, but the meaning of each part is unknown. The meaning of object parts is crucial for applying textures and materials. For example, it’s common that door panels are made of wood and door knobs are made of metal, but we can’t assign correct materials to them if we don’t know which part is which. In order to support our scene generation pipeline, especially the style compatibility, we manually label the meaning of each part for every model. Then, a common material is assigned to each object part. As shown in Figure \ref{model-part-labeling}, three objects are labeled with part meaning and material.



 \begin{figure}[ht]
 	\centering
 	\includegraphics[width=3.0in]{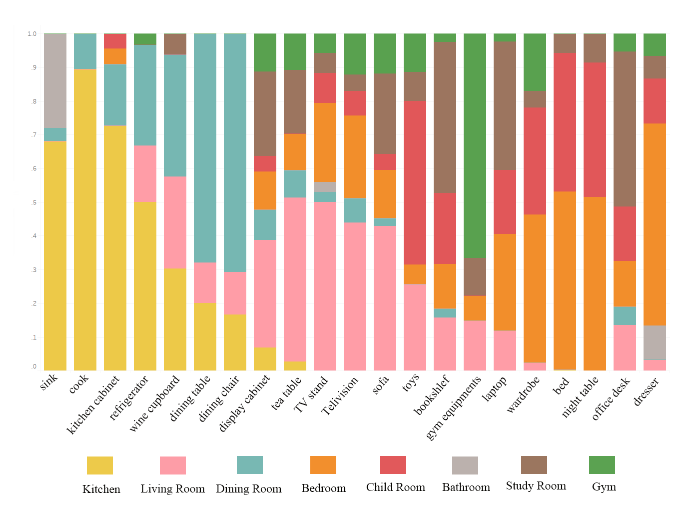}
 	\caption{Degree of relevance between room type and object category.}
 	\label{roomtype-obj}
 \end{figure}

\section{Approach}
\subsection{Knowledge Building}  \label{exp}
In this paper, we adopt L-LDA to recommend indoor room content based on specific room type. To evaluate the recommend instance category lists, we conduct user study to build up a knowledge of relationship between room type and instance category. In total, 8 room types and 30 basic interior instance categories are selected. We randomly invite 30 subjects from societies and they were told to score relationship between room type and instance categories. Partial results are illustrated in Figure \ref{roomtype-obj}. For a given room which contains a list of instance categories $O= ( o_{1},\cdots ,o_{N})$, user study gives a score $S= ( s_{1},\cdots ,s_{N})$ correspondingly. The match expectation of the given list and the room type can be calculated by:
$$E\left ( O,R_{T} \right )=\sum_{i=1}^{N}p\left ( o_{i} \right )\cdot s_{i}$$
where $p(o_{i})$ is the presence probability of $o_{i}$.

\subsection{Category-level recommendation}\label{cl}
\begin{figure}[ht]
  \centering
  \includegraphics[width=3in]{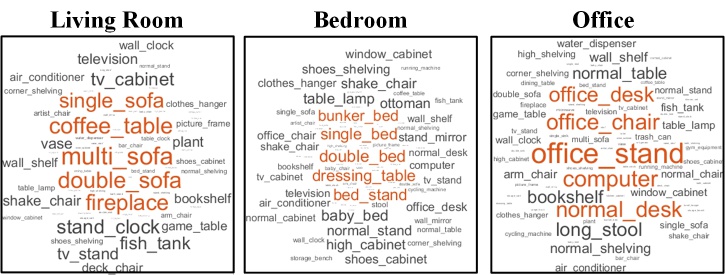}
  \caption{Three example recommendation lists learned from S-SUNCG by L-LDA.}
  \label{category-recommendation}
 \end{figure}

\begin{figure}[ht]
  \centering
   \includegraphics[width=3in]{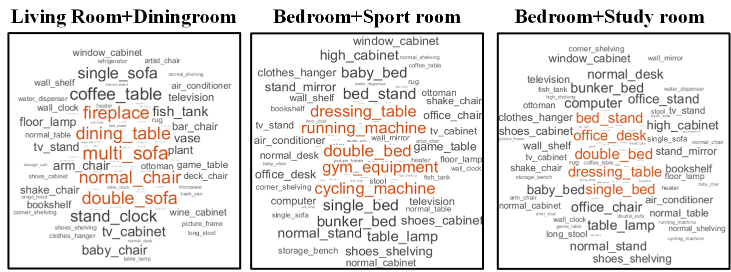}
   \caption{Recommendation lists of three hybrid rooms.}
   \label{hybrid room}
\end{figure}

We use the single-type rooms from S-SUNCG as training data, and train L-LDA \cite{ramage2009labeled} model with the semantic information of object categories in each room. As described in Sec.\ref{room type}, $p(o_{i}|R_{T}^{j})$ indicates the probability that instance category $i$ exists in the room with type $j$. Figure \ref{category-recommendation} shows three examples which are living room, bedroom and office. According to joint probability of room type and instance category, the occurance probability of instance categories in hybrid room can be calculated by: 
$$p\left ( o_{i} | R_{T}^{j},R_{T}^{k}\right )=p\left ( o_{i}|R_{T}^{j} \right )+p\left ( o_{i}|R_{T}^{k} \right )$$
where $(R_{T}^{j},R_{T}^{k})$ are the types of hybrid room. Figure \ref{hybrid room} shows three examples of hybrid rooms.

\subsubsection{List similarity}

\begin{figure}[ht]
  \centering
  \includegraphics[width=3in]{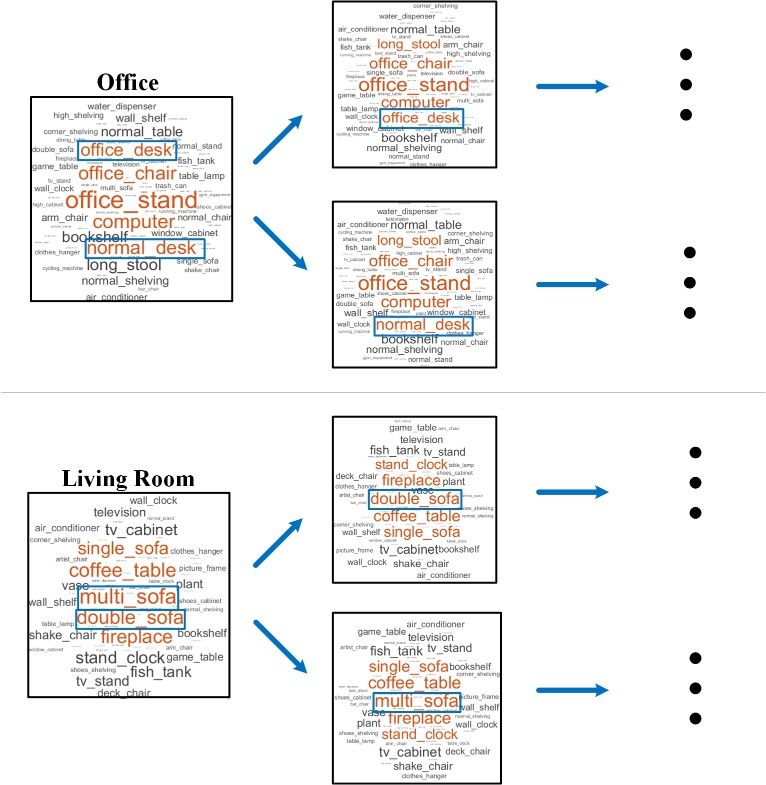}
  \caption{Recommendation list spilt by substitutional relation check.}
  \label{list-split}
\end{figure}

Since L-LDA is a probabilistic model to generate a instance recommendation list, which means some instance categories have a high probability of appearance, but when making recommendations, these categories cannot appear in the same room. For example, the categories of single bed and double bed can appear in the same bedroom with remote possibility, although they both have high appearance probabilities. Another two examples are shown in Figure \ref{list-split}, the two pairs of categories marked in blue boxes present substitutional relations in each recommendation list. To solve this relation and increase the optional richness of recommendation lists, we use a paradigmatic model \cite{pennington2014glove} to calculate the similarity of two categories, and split the lists according to the substitutional relation check. For a list with few categories $O= ( o_{1},\cdots ,o_{N})$, we first generate its co-occurrence matrix $C$ by statistics. The category similarity can be calculated by:
$$S=\sum_{i,j=1}^{N}f(C_{ij})\left ( O_{i}^{T}O_{j}+\xi _{i}+\xi _{j}-log\left ( C_{ij} \right ) \right )^{2}$$
where $\xi _{i}$ and $\xi _{j}$ are two offset items, $f\left ( \cdot  \right )$ is a weight item based on word frequency:
$$f\left (x \right )=\left\{\begin{matrix}
\left ( x/x_{max} \right )^{\alpha } \, \, \, \, \, \, \, \, & if  x< x_{max}\\ 
 1\, \, \, \, \, \, \, \, & otherwise
\end{matrix}\right.$$
As mentioned in \cite{pennington2014glove}, $f\left ( \cdot  \right )$ works best when $\alpha=3/4$. We assign the similarity threshold to 0.95, and perform pairwise substitutional relation check by iterating . The second column in Figure \ref{list-split} shows the split lists after first iteration. 

 \begin{figure*}[ht]
  \centering
  \includegraphics[width=6in]{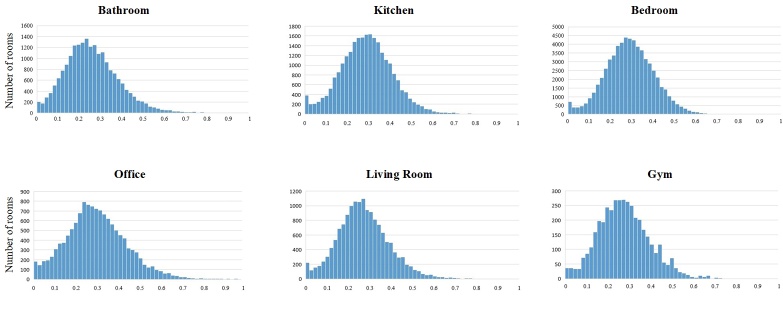}
  \caption{The statistic occupation proportion distributions of six room types in S-SUNCG.}
  \label{room-size}
\end{figure*}

\subsubsection{List recommendation}

According to the annotated floor plan, a house can be represented as a set of instance category lists corresponding to the room type, which is $H=\left \{L_{1},\dots,L_{N} \right \}$. Among them, the hybrid type rooms are represented by a combination of single type lists. We divide the lists into two parts: lists appear only once in the house and lists appear more than once. Recommend the most suitable list for the house through combinatorial optimization. For the first part, we calculate the match expectation (Sec.\ref{exp}) as: 
$$H_{1}= \sum_{i=1}^{n}E(L_{i},R_{T}^{i})$$
where $L_{i}$ indicates only one room of this type in the house, $R_{T}^{i}$ is the corresponding room type and  $E(\cdot)$ is the match expectation. The second part is calculated as:
 $$H_{2}=\sum_{i=1}^{m_{i}}\sum_{j=1}^{k}E\left ( L_{i},R_{T}^{j} \right )$$
 where $k$ is the number of room types which means two more rooms exist in the house, $m_{i}$ indicates the number of rooms belong to the same type. The instance category recommendation lists of the entire house can be selected by:
 $$H=\mathop{\arg\max}_{L} (H_{1}+\lambda H_{2})$$
 where $\lambda$ is a weight to control the dissimilarity of recommendation lists for rooms of the same type. we assign $\lambda=1$ to keep the dissimilarity. The lists of hybrid rooms can be generated by combination of single type lists as described in Sec.\ref{cl}.

\subsection{Instance-level recommendation}
 We cast the problem of instance selection as a combinatorial optimization. For a specific house with a set of selected recommendation lists, our instance suggestion is to find a combination of instances for the entire house from the lists, which best satisfies both size compatibility and style compatibility.

\subsubsection{Occupation constraint}\label{531}

A specific size room can accommodate limited number of furniture considering the needs of human activities. Thus, we define a object occupation proportion for a room as follows:
$$g_{1}(o,r)=\frac{\sum_{i}^{n_{c}}\vartheta _{i}S(o_{Mask}^{i})}{S(r_{Mask})}$$
where $n_{c}$ is the number of categories in corresponding recommendation list of the room, and $S(\cdot)$ is the area function. $o_{Mask}$ and $r_{Mask}$ represent respective projections on the ground. Here we only calculate the masks of objects which are on the floor, and the objects like rugs which do not affect human activities are excluded. $\theta_{i}$ indicates the number of instance which belongs to $i$th category,
$$\theta _{i,j}\in \left\{\begin{matrix}
\left [0,  n_{i,j}\right ] \, \, \, \, \, \, \, \,   & C_{i,j}\leq 2.5\\ 
\left [1,  n_{i,j}\right ] \, \, \, \, \, \, \, \,  & otherwise
\end{matrix}\right.$$
where $C_{i,j}$ is the score of $i$th category in $j$th room type from user study (Sec.\ref{exp}). $n_{i,j}$ is the corresponding prior value learn from S-SUNCG dataset.
\begin{figure*}[t]
  \centering
  \includegraphics[width=6in]{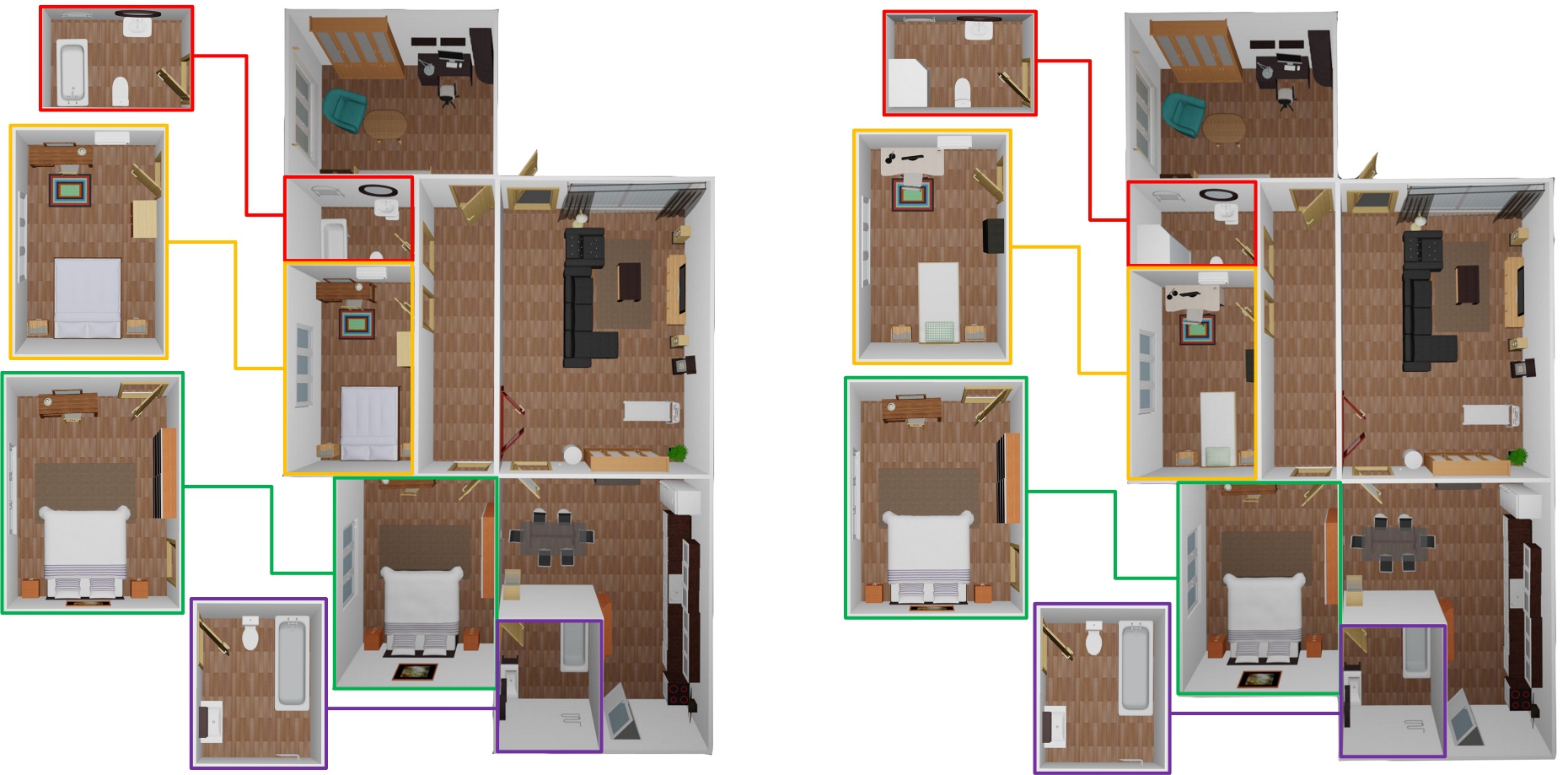}
  \caption{The effect of dissimilarity control weight:  left $\lambda=0$ versus right $\lambda\neq 0$.}
  \label{dissimilarity-control-term}
\end{figure*}

\subsubsection{Style constraint}\label{532}

We use a stylistic image as input to constrain instance suggestion. Inspired by \cite{chen2015magic}, a local group rule and a globe aesthetic rule are built.

\paragraph{Group rule} In the real world indoor scene, instances in the same room share same material, although they belong to different categories (e.g night stand and bed frame have similar material). As described in \cite{chen2015magic}, a binary likelihood is defined to account for this property. However, this is a strong constraint when applied from single room to entire house, a large number of furniture parts select same material. In our instance suggestion, we adopt CSR test \cite{shaokui} to build local instance groups for each room. $t(o_{1},o_{2})$ is defined as a CSR test, value $1$ if the two instances pass the CSR test, value $0$ otherwise. For a given room, local groups can be obtained by CSR tests. In each group, binary likelihood is used to constrain instance selection. We use the same notation as mentioned in \cite{chen2015magic}, a material is represented as $m$. In a specific instance group which pass the CSR test, we define the group rule as follows,
\begin{equation*}
\begin{split}
&g_{2}(o_{1},c_{1},o_{2},c_{2})= \\
 &t(o_{1},o_{2}) \sum_{{m}_{1}',{m}_{2}'}\frac{n(o_{1},o_{2},c_{1},c_{2},m_{1}',m_{2}')}{n(o_{1},o_{2},c_{1},c_{2})} \prod_{i=1,2}s(m_{i},m_{i}')\\
 \end{split}
 \end{equation*}
 where $o_{1}$, $o_{2}$ are two instances in the group, $c_{1},m_{1}$ and $c_{2},m_{2}$ indicate corresponding part categories and materials. $n(o_{1},o_{2},c_{1},c_{2},m_{1}',m_{2}')$ is the number of time that two instance parts co-occur $(m_{1}',m_{2}')$, and $n(o_{1},o_{2},c_{1},c_{2})$ is the total number of two instance parts co-occur any set of materials. $s(\cdot)$ indicates material similarity function, value $1$ or $0$ refer to similar or not.  

 \paragraph{Global style rule} A five-color palette is used to represent the color theme of the input picture \cite{chen2015magic}. $M(p,{p}')$ represents the similarity of two palettes:
$$M(p,{p}')=\sum_{i=1}^{5}\sum_{j\in\left \{ h,s,v \right \} }w_{j}(p_{i,j}-p_{i,j}')^{2}$$
where $p_{i,j}$ indicates the $j$th channel of $i$th palette in HSV color coordinate system. $w_{j}$ is the weight for each channel, we set $w_{h}=w_{s}=1$, $w_{v}=3$ as mentioned in \cite{chen2015magic}. To generate color palette for hybrid rooms, $D(p,{p}')$ is defined to calculate the hybrid color palette of $p$ and ${p}'$,
$$D(p,{p}')=\sqrt{\frac{p_{i,j}^{2}+{p}_{i,j}'^{2}}{2}}|_{i\in \left [ 1,5 \right ],j\in\left \{ h,s,v \right \}}$$

\subsubsection{Cost function}
Given an annotated house, in which each room involved a instance category recommendation list. Our cost function for instance selection includes three items: an occupation constraint item $G_{1}$, a local group constraint item $G_{2}$, a global style constraint item $G_{3}$:
$$G(O)=\mu_{1}G_{1}(O,R)+\mu_{2}G_{2}(O,C)+\mu_{3}G_{3}(O)$$
where $\mu_{1}$, $\mu_{2}$ and $\mu_{3}$ are weights to balance the three items. The sensitivity of optimization equations to coefficients is not high, therefore, we set $\mu_{1}=\mu_{2}=\mu_{3}=1/3$ in our experiments. 

\paragraph{Occupation constraint item} The occupation constraint item considers how suitable the total occupation area proportion of selected instances in the entire house. It is defined according to the occupation constraint in Sec.\ref{531},
$$G_{1}(O)=\frac{1}{2n_{r}}\sum_{i=1}^{n_{r}}(g_{1}(o_{i},r_{i})-\delta _{i})^{2}$$
where $n_{r}$ indicates the total number of rooms in the processing house. $\delta_{i}$ is the optimal proportion of total instance occupation area in $i$th room type which is learn from dataset. Figure \ref{room-size} shows the statistic occupation proportion distributions of six room types in S-SUNCG. The optimal proportions of each room type can be obtained by Gaussian fitting.

\paragraph{Group constraint item} The group constraint item considers material compatibility between instance parts in the groups. It is defined according to the local group rule for all groups in the entire house,
$$G_{2}(O,C)=-\sum_{k=1}^{n_{g}}\sum_{1\leq i< j\leq n_{k}}\frac{2\cdot log(g_{2}(o_{i},c_{i},o_{j},c_{j})+\xi )}{n_{k}(n_{k}-1)}$$
where $n_{g}$ indicates the total number of local groups in the house, $n_{k}$ is number of instances in the corresponding group. $\xi$ takes $1e-5$ to avoid $0$ value.

\paragraph{Global style constraint item} The global style constraint item considers style compatibility of entire house based on the input stylistic image. The item can be defined as:
$$G_{3}(O)=e^{-\sum_{i=1}^{n_{r}}M(p_{t}^{i}, p_{i}(o))}$$
where $n_{r}$ indicates the total number of rooms. $p_{i}(o)$ indicates the extracted five-color palette of $i$th room. $M(\cdot)$ is the similarity of two palettes as defined in Sec.\ref{532}. To maintain the style compatibility of the entire house, we search a house style template from Structured3D dataset according to the input stylistic image. Since the user only input one stylistic image for one room, to maintain the style compatibility of entire house, we search the most similar picture of same room type in the Structured3D \cite{zheng2019structured3d}, and map the pictures from other types of rooms to our processing house. Involving the hybrid rooms without style reference images, we generate color palettes from the style reference images of single type rooms as described in Sec.\ref{532}.
\begin{figure*}
  \centering
  \includegraphics[width=6in]{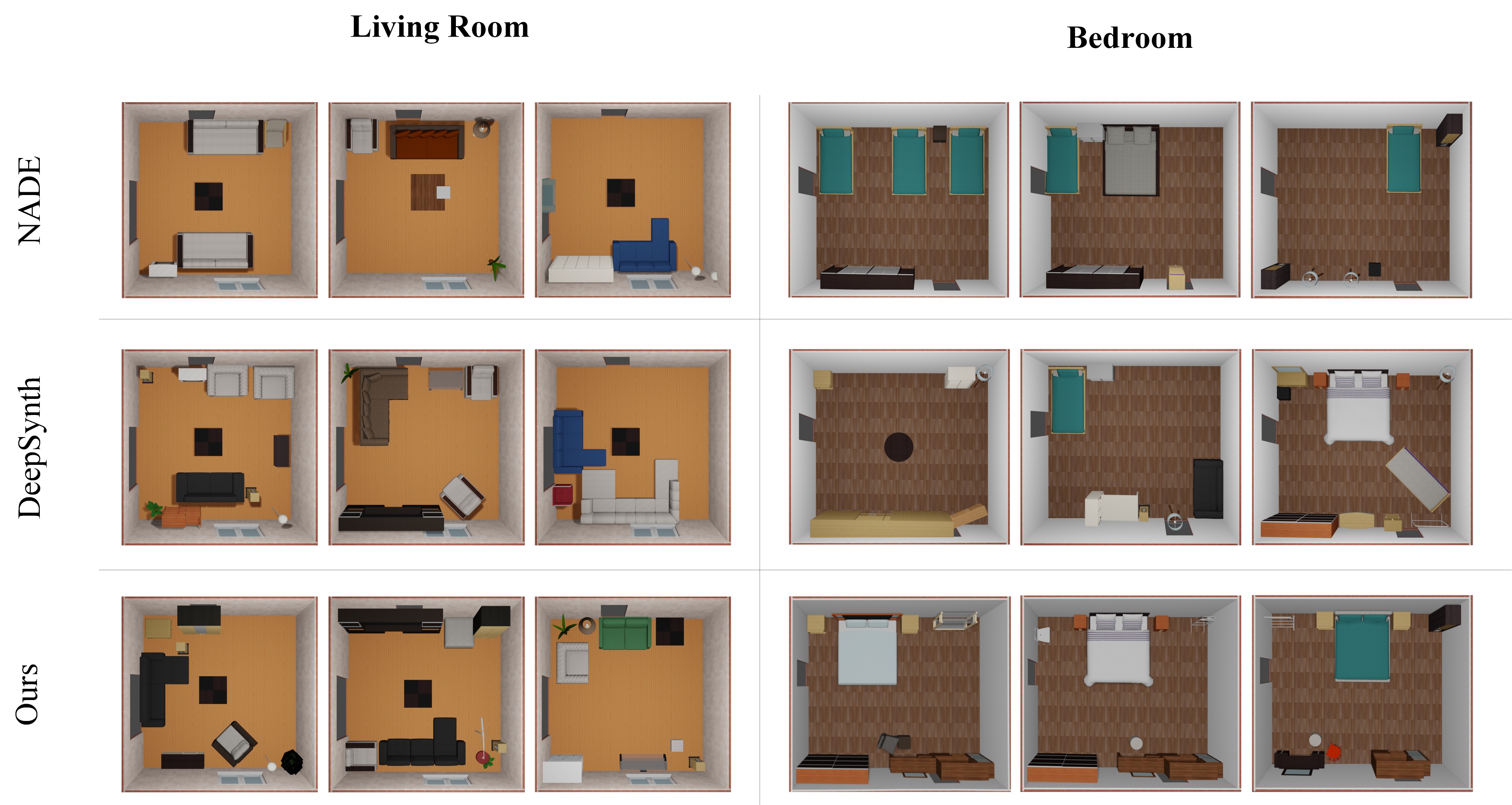}
  \caption{Comparison results on category-level recommendation with other two baseline methods.}
  \label{comparison-selection}
\end{figure*}

\section{Evaluation and Results}
\subsection{Effects of control weights in category-level recommendation}

Figure \ref{dissimilarity-control-term} evaluates the effect of dissimilarity control weight $\lambda$ in the category list recommendation step. As shown in Figure \ref{dissimilarity-control-term} (Right), when $\lambda\neq0$, the rooms of same type will select different recommendation lists. The bedroom in green uses a list with a double bed and a dressing table, while the bedroom in yellow uses a list with a single bed and an office desk. The bathroom in red uses a list with a shower, while the bathroom in purple uses a list with a bathtub. Without the dissimilarity control weight, the bedrooms and bedrooms most likely end up using the same lists, as shown in Figure \ref{dissimilarity-control-term} (Left). For the entire house, the object combination in Figure \ref{dissimilarity-control-term} (Right) can satisfy richer functional requirements.

\begin{figure}[h]
  \centering
  \includegraphics[width=3in]{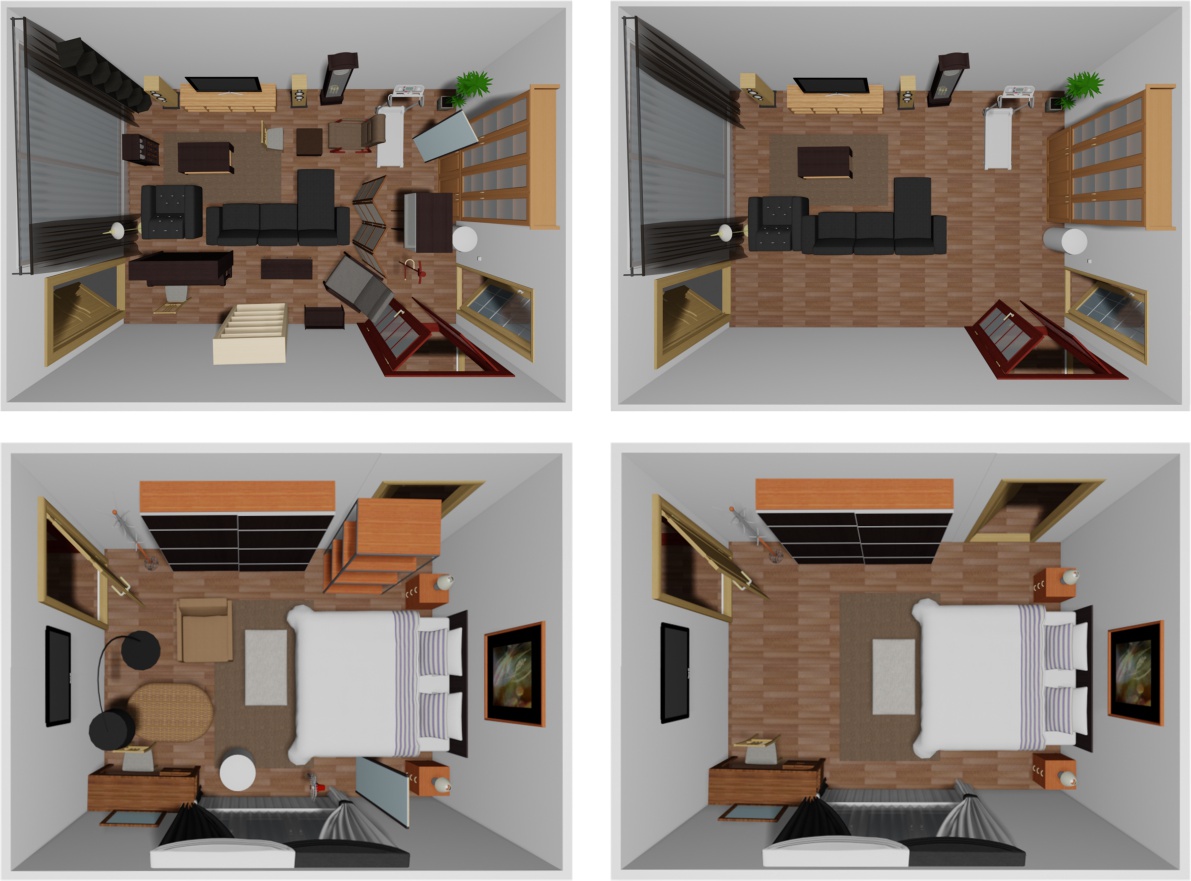}
  \caption{Effect of occupation constraint item, $\mu_{1}=0$ (left) versus $\mu_{1}\neq 0$ (right). }
  \label{size-constraint}
\end{figure}

\subsection{Comparison on category-level recommendation}

To show the effectiveness of our topic model on category-level recommendation, we compare our method to some baselines, namely Neural Autoregressive Distribution Estimator (NADE) \cite{uria2016neural} and Deep Convolutional Priors (DeepSynth) \cite{wang2018deep}. NADE models the probability distribution over a sequence of variables, while DeepSynth utilizes convolutional neural networks (CNNs) for occurrence modeling. For both NADE and DeepSynth, we adopt the training strategy described in \cite{wang2018deep}. We apply our method to the original SUNCG object categories for fair comparison. For NADE and our method, we first get a sample of occurrence counts of all categories, then synthesize scenes forcing DeepSynth to select instances, and finally arrange them according to the occurrence counts.

Figure \ref{comparison-selection} shows several representative examples generated by each method. Note that we evaluate the results for two different types of rooms, i.e., bedroom and office, and the geometry for each type of room is the same. We can see from the results that our method 
performs category-level recommendation with proper object functions and reasonable object counts. In contrast, NADE leads to multiple beds in the bedroom and DeepSynth generates results less diversity of object function. 

\begin{table}%
\caption{User study results of comparison on category-level recommendation. Users are shown results generated by three methods, order of which is randomized, and are asked to rank the three results from the perspective of object rationality and function diversity. Rank 1,2,3 gets 1,3,5 points respectively. This table reports average scores of all participants.}
\label{comparison-selection-table}
\begin{minipage}{\columnwidth}
\begin{center}
\begin{tabular}{ccc}
  \toprule
  Method & Living Room & Bedroom\\ \midrule
  NADE \shortcite{uria2016neural}     & 1.2 & 1.7\\
  DeepSynth \shortcite{wang2018deep}  & 2.8 & 3.4\\
  Ours     & 4.5 & 3.9\\
  \bottomrule
\end{tabular}
\end{center}
\end{minipage}
\end{table}%

\begin{figure}
  \centering
  \includegraphics[width=3in]{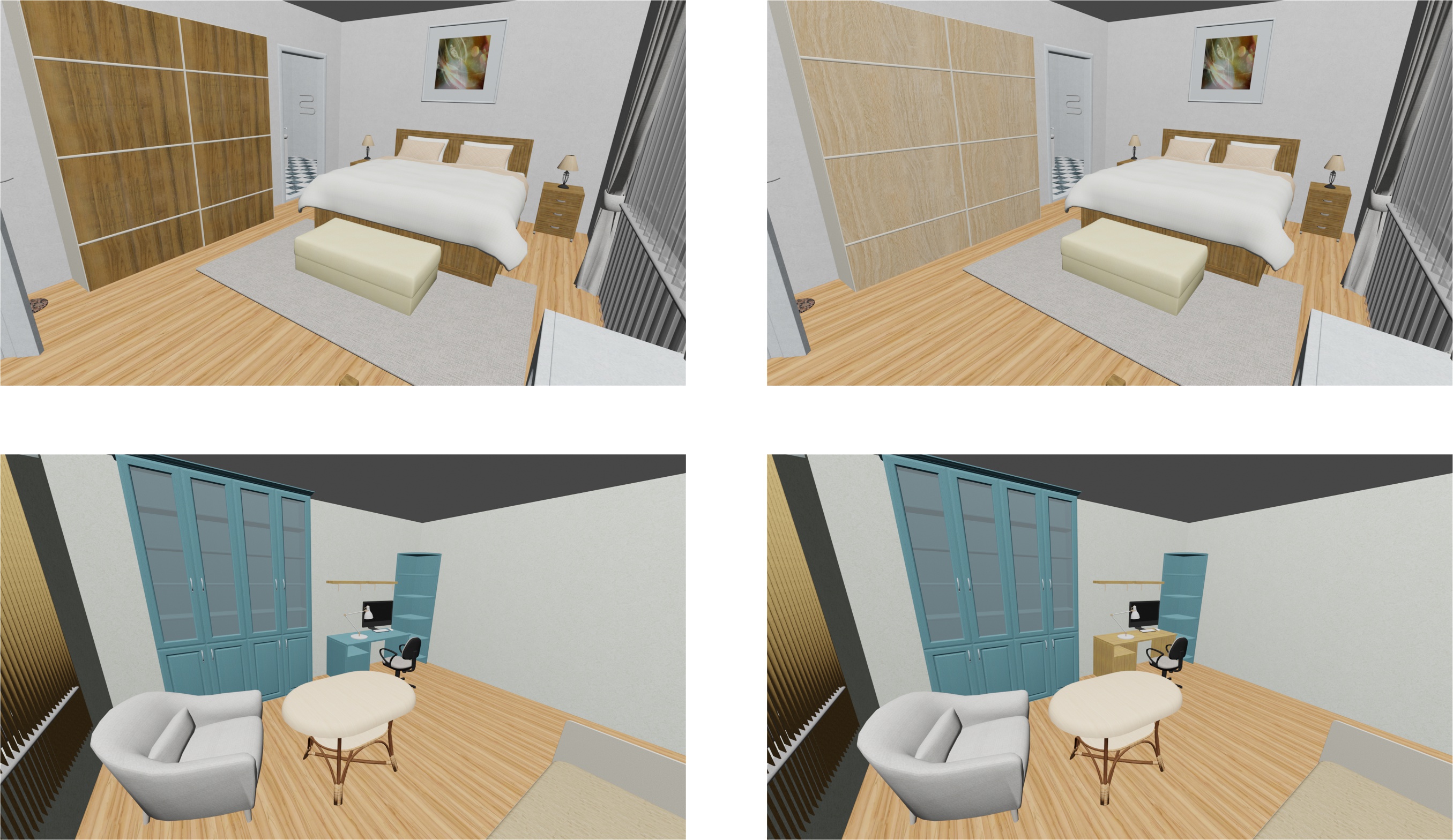}
  \caption{Comparison results of group constraint with baseline method. The left column adopts binary likelihood of \cite{chen2015magic}, and the right column adopts group constraint. }
  \label{group-vs}
\end{figure}

\begin{figure*}[h]
  \centering
  \includegraphics[width=6in]{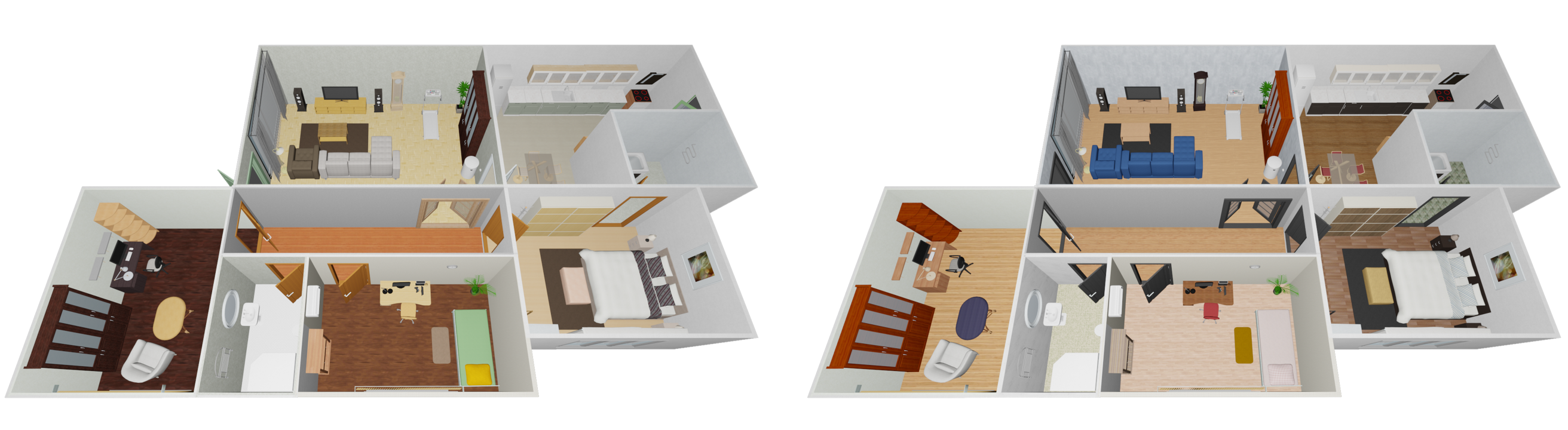}
  \caption{Effect of group constraint item, $\mu_{2}=0$ (left) versus $\mu_{2}\neq 0$ (right). }
  \label{group-item}
\end{figure*}

\subsection{Effects of energy terms in instance-level recommendation}
Figure \ref{size-constraint} shows the effect of occupation constraint, group constraint and style constraint terms of our cost function in instance recommendation step. We fix the materials of instances in the evaluation of occupation constraint term. As illustrated in left column of Figure \ref{size-constraint}, without occupation constraint term, each room will continuously sample instances until the recommendation list is traversed or there's no more space for a new instance. Because of that, the synthesized rooms will look very crowded and chaotic. The right column of Figure \ref{size-constraint} shows better results with the occupation constraint term.

\begin{figure*}[h]
  \centering
  \includegraphics[width=6in]{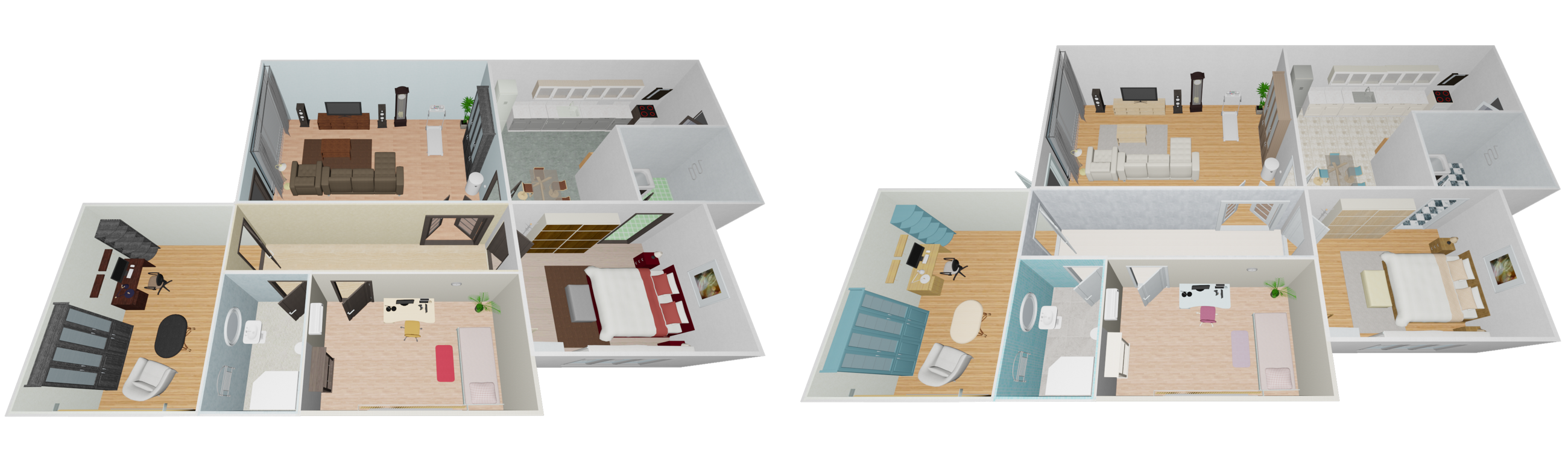}
  \caption{Effect of global style constraint item, $\mu_{3}=0$ (left) versus $\mu_{3}\neq 0$ (right). }
  \label{global-item}
\end{figure*}

To evaluate the other two terms, we select a specific set of instances for experiments. The left side of Figure \ref{group-item} shows obvious material inconsistency between associated objects (e.g beds and night stands, TV stand and coffee table), which are generated ignoring the group constraint term. However, such object groups with specific function usually maintains material consistency in reality. The right side of Figure \ref{group-item} shows the material consistency in the object groups considering the group constraint term. The synthesized scene looks more harmonious in rooms with more objects groups. Compared to the binary term in \cite{chen2015magic}, material consistency between objects can be more accurately constrained through our group constraint term. As is shown in the left side of Figure \ref{group-vs}, the constraint of \cite{chen2015magic} is too strong, so that a large number of objects in the same room share the same material. When dealing with the external style template constraints as mentioned in this article, \cite{chen2015magic} cannot give satisfactory results. Figure \ref{global-item} evaluates the global style constraint term, as is shown in the right side of Figure \ref{global-item}, the object selection of each room meets the style template of the corresponding reference image, and the combination of multiple rooms meets the style of the entire house, which appears to be more style compatible than the right side of Figure \ref{global-item}. Thus, the above three constraints are essential for our instance recommendation.

\subsection{User Study}
In order to verify the effectiveness of our method, which reflected by the rationality and aesthetics of the scenes produced by the pipeline, we invited 25 subjects to participate in the user study, 20 of whom are ordinary users and 5 are designers. 

In our user study, we render five pairs of images for each participant. In each pair, one is the original scene from the SUNCG dataset, and the other one generated by our pipeline. These two scenes use the same floorplan. Each pair of images are rendered with the same light condition in the same view, and is presented in a random order. 

\begin{figure*}[h]
  \centering
  \includegraphics[width=6in]{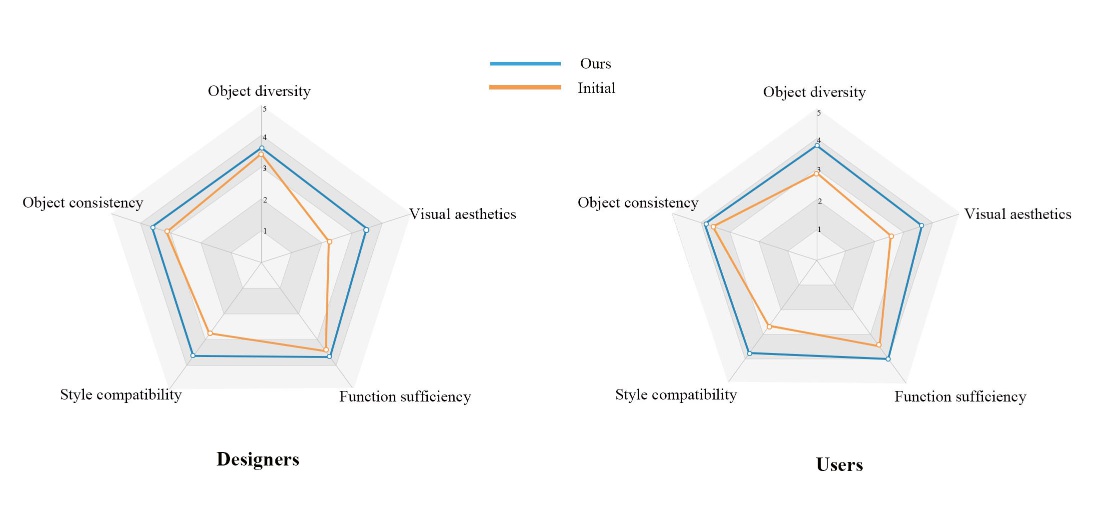}
  \caption{Average user rating scores of the original S-SUNCG scene and the scene synthesized by our approach in five different aspects.}
  \label{user study}
\end{figure*}

We asked the subjects to rate each image from 0 (worst) to 5 (best) in five aspects, namely object diversity, object consistency, style compatibility, function sufficiency and visual aesthetics of the scene. Figure \ref{user study} plots the average score of each item rated by users and designers, which shows that the scenes produced by our pipeline get much higher scores than the initial scenes designed by designers in all five aspects, especially in style compatibility and visual aesthetics. In that case, it is verified our pipeline is of good efficiency in producing plausible scenes comparable to those produced by designers.

\section{Conclusions}

In this paper we presented a novel approach to synthesize style-compatible indoor scenes for multiple rooms in a residence. For a practical concern, we assume a room can have multiple room types to allow objects with different functions. We focus on object recommendation and achieve it from category selection to instance selection. We consider function diversity and style compatibility in both room-level and residence-level. Experiments show that our approach generates high-quality multi-room indoor scenes comparable to those produced by professional interior designers.

There are several limitations of our work for future improvement. For example, style compatibility is implemented based on a 5-color palette of the reference image, while it is more reasonable to extract textures directly from objects presented in the image. On the other hand, our work is more restricted to color style but shape consistency should also be considered during object selection.

%
%
%
%

\bibliographystyle{ACM-Reference-Format}
\bibliography{sample-bibliography}

\end{document}